%% file: PoS_Lat14.tex
\documentclass[cits]{PoS}

\title{On the extraction of spectral quantities with open boundary conditions}

\ShortTitle{On the extraction of spectral quantities with open boundary conditions}

\author{\speaker{Mattia Bruno} $^a$,  Piotr Korcyl$^a$, Tomasz Korzec$^b$, Stefano Lottini$^a$ and Stefan Schaefer$^a$\\
\llap{$^a$} John von Neumann Institute for Computing (NIC), DESY\\
Platanenallee 6, D-15738 Zeuthen, Germany\\
\llap{$^b$} Institut f\"ur Physik, Humboldt Universit\"at zu Berlin\\
Newtonstr. 15, D-12489 Berlin, Germany}

\newcommand{\preprintline}{\vspace{3cm}\newline
\rightline{\parbox{2.9cm}{
\large\tt DESY 14-211
\large\tt HU-EP-14/50}}
}

\abstract{We discuss methods to extract decay constants, 
meson masses and gluonic observables in the presence of open boundary conditions. 
The ensembles have been generated by the CLS effort and
have 2+1 flavors of O(a)-improved Wilson fermions 
with a small twisted-mass term
as proposed by L\"uscher and Palombi. 
We analyse the effect of the associated reweighting factors on the computation
of different observables.
\preprintline}

\FullConference{The 32nd International Symposium on Lattice Field Theory,\\
		23-28 June, 2014\\
		Columbia University New York, NY}

\newcommand{\meff}{m_\mathrm{eff}}

\pdfoutput=1   % needed on arXiv if pdflatex is used

\begin{document}

\input{pos_text}

\bibliography{PoS2014}
\bibliographystyle{JHEP}

\end{document}

%% file: pos_text.tex
\section{Introduction}

To overcome the topological freezing which makes 
simulations at lattice spacings below 0.05~fm practically impossible, 
open boundary conditions in the temporal direction 
have been proposed~\cite{Luscher:2011kk}.
In our study we analyse the ensembles
generated by the CLS effort with 2+1 $\mathrm{O}(a)$-improved Wilson
fermions and twisted-mass reweighting \`a la 
L\"uscher-Palombi \cite{Luscher:2008tw}.
For the details of the algorithmic and physical 
parameters we refer to Refs.~\cite{PiotrPoS,CLSpaper}.
In these Proceedings we discuss the consequences of 
the open boundaries in 
gluonic and fermionic observables, such as the scale parameter $t_0$~\cite{Luscher:2010iy}
and the pion mass and decay constant.
In the last section we also examine the implications of the 
twisted-mass reweighting and its impact on the final results.

\section{Definitions}
The interpretation of the boundary state can be taken 
from Ref.~\cite{Guagnelli:1999zf}, where Schr\"odinger-functional 
boundary conditions are adopted.
Our case is similar, since far away from the boundary 
the expectation value of an observable $O(x_0, \mathbf{x})$
corresponds to the vacuum expectation value up to 
exponentially suppressed contributions from 
states with vacuum quantum numbers (e.g. a $2 \pi$ state in QCD or a scalar glueball
 in pure Yang-Mills theory).
The gluonic observable under study is the energy density (clover-type discretisation of $\hat{G}_{\mu\nu}$) 
obtained from the Wilson flow at positive flow time~\cite{Luscher:2010iy}
\begin{equation}
t^2 \langle E(t,x_0) \rangle = t^2 \frac{1}{L^3} \sum_\mathbf{x} \frac{1}{4} 
\langle \hat{G}_{\mu \nu}^a(x_0,\mathbf{x}) \hat{G}_{\mu \nu}^a(x_0,\mathbf{x}) \rangle \,,
\label{eq:energy_density}
\end{equation}
which we also use to compute the scale $t_0$, defined by the time $t$ at which 
the r.h.s of eq.~(\ref{eq:energy_density}) is equal to 0.3.
From the fermionic side, we measure the following two-point correlation 
functions stochastically\footnote{The measurement 
code is publicly available at  \texttt{https://github.com/to-ko/mesons}.}
(throughout the paper $T$ and $L$ refer to the temporal and spatial extent of the lattice)
\begin{equation}
f_X(x_0,y_0) = - \frac{a^6}{L^3} \sum_{\bf x,y} 
\langle X^{rs}(x_0,{\bf x}) P^{sr}(y_0,{\bf y}) \rangle \,, \quad
P^{rs}(y_0,\mathbf{y}) = \overline{\psi}^r (y) \gamma_5 \psi^s (y) \,, 
\label{eq:fP_fA}
\end{equation}
with $X^{rs}$ either $P^{rs}$ or $A_0^{rs}(y_0,\mathbf{y}) = 
\overline{\psi}^r (y) \gamma_0 \gamma_5 \psi^s(y)$ and $r,s$ flavor indices. 
We compute the effective mass using
\begin{equation}
a \meff(x_0 + a/2,y_0) = \log \frac{f_\mathrm{P} (x_0,y_0)}{f_\mathrm{P} (x_0+a,y_0)} \,, \quad 
\meff^\mathrm{aver}(y_0) = \frac{1}{N_\mathrm{points}} \sum_{x_0 \in \mathrm{plateau}} \meff(x_0,y_0) \,.
\label{eq:meff}
\end{equation} 

\section{Cutoff effects}

\begin{figure}
\includegraphics[width=\textwidth]{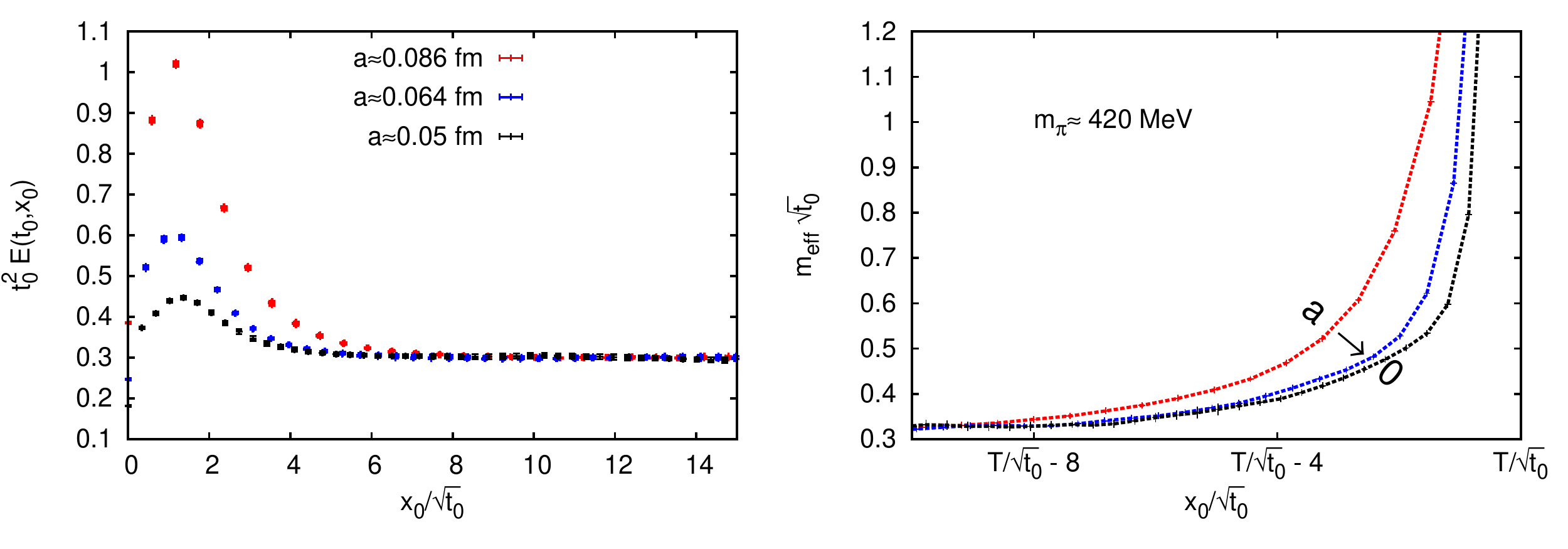}
\caption{\textit{Left}: cutoff effects of the energy density in units of $t_0$ 
close to the boundary at $x_0=0$ for three lattice spacings and
pion masses from 200~MeV to 420~MeV.
\textit{Right}: cutoff effects of $\meff(x_0 + a/2,y_0=a)$ close to the boundary at $x_0=T$ in units of $\sqrt{t_0}$.}
\label{fig:boundary}
\end{figure}

In Figure \ref{fig:boundary} we show the profile of the energy density and the effective pion mass as the boundaries
at $x_0=0$ and $x_0=T$ are respectively approached. 
In $E(t_0,x_0)$ the effects seem to be dominantly of $\mathrm{O}(a)$, 
indeed the plateau starts (for our three lattice spacings) always in the same region of $x_0/a$.
Moreover effects from the sea-quark masses are not visible, even at the boundaries, 
given the sub-percent precision of this observable.
In the pion mass, sea-quark effects close to the boundaries are present, 
though much smaller than the dependence on the spacing.

As a measure of the goodness of our plateaux we fit both quantities
to a constant and we compare the uncorrelated $\chi^2$
to $\chi^2_\mathrm{exp}$, the expected value computed from
the measured covariance matrix\footnote{This 
method avoids the inversion of the covariance matrix.},
as a function of the distance from the boundary.
We find that in the central part of the lattice 
$\chi^2/\chi^2_\mathrm{exp}$ is around one.
For the energy density we take the first time slice where
$\chi^2/\chi^2_\mathrm{exp} \approx 1$ as the beginning 
of the plateau, while for the effective mass eq.~(\ref{eq:meff}) 
we consider an additional distance of 0.25~fm 
to avoid residual excited states
(for a multi-state analysis see Ref.~\cite{CLSpaper}).

\section{Decay constants}

As a first attempt, one would naively place source and sink of the two-point functions 
in the middle of the lattice, to avoid boundary contaminations.
However the additional contributions of the states excited 
by the source operator would reduce even more the portion
of the lattice usable for the extraction of a mass or a decay constant. For this reason we measure the 
correlators with the source close to one boundary and the sink in the bulk. When possible we always make
 use of time reversal symmetry by averaging correlation functions at $(x_0,y_0)$ and $(T-x_0,T-y_0)$.

Now, the expectation values in eq.~(\ref{eq:fP_fA}) take the form (for sufficiently large $x_0-y_0$ and $T-x_0$)
\begin{equation}
f_\mathrm{A} (x_0,y_0) = A(y_0) \hat{f}_\pi e^{-\meff (x_0 - y_0)} \,, \quad f_\mathrm{P} (T-y_0,y_0) = A(y_0)^2 e^{-\meff (T - 2 y_0)} \,.
\label{eq:fA_ansatz}
\end{equation}
Note that $\hat{f}_\pi$ indicates the matrix element $\langle 0 |\mathbb{A}_0|\pi \rangle$ related to the pion decay constant through an additional normalisation factor. 
$A(y_0)$ is the amplitude related to the matrix element $\langle 0|\mathbb{P}|\pi \rangle$
 which depends on the distance from the boundary $y_0$
 because we place $y_0$ close to one boundary where the excited states from the boundary 
are not yet exponentially suppressed.

In Ref. \cite{Guagnelli:1999zf}, where the same type of observables was studied in the Schr\"odinger functional setup, 
the following ratio to compute
the pion decay constant has been proposed
\begin{equation}
F_\pi^\mathrm{bare} \propto \frac{f_\mathrm{A}(x_0,y_0)}{\sqrt{f_\mathrm{P}(T-y_0,y_0)}} e^{\meff (x_0 - T/2)} \,.
\label{eq:fpi_guagnelli}
\end{equation}
Alternatively we consider here a new combination of two-point functions which cancels at the same time the amplitude $A(y_0)$ and
the remaining exponential factor
\begin{equation}
F_\pi^\mathrm{bare}(x_0,y_0) \propto 
\sqrt{\frac{f_\mathrm{A}(x_0,y_0) f_\mathrm{A}(x_0,T-y_0)}{f_\mathrm{P}(T-y_0,y_0)}} \,.
\label{eq:fpi_method}
\end{equation}
In both cases a vacuum average in the plateau region $0 \ll x_0 \ll T$ can be taken.

\begin{figure}
\includegraphics[width=\textwidth]{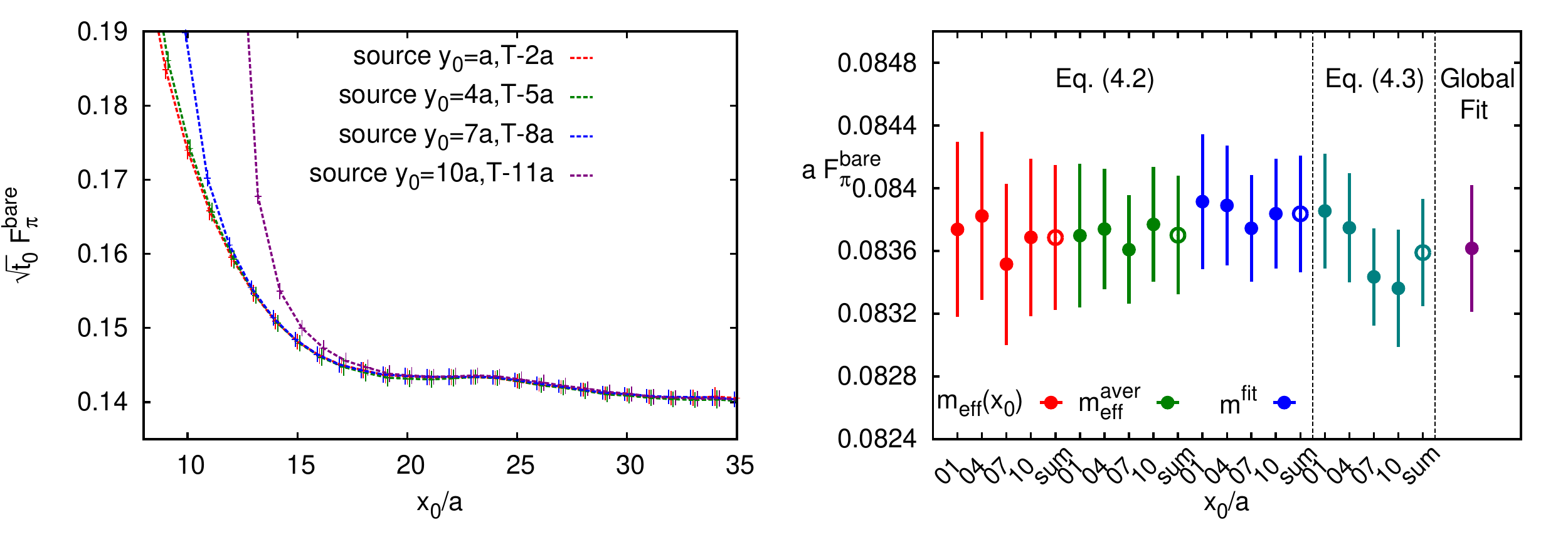}
\caption{The left panel shows the pion decay constant computed from eq.~(\protect\ref{eq:fpi_method}) for several
source positions. In the right panel we plot the results from eq.~(\protect\ref{eq:fpi_guagnelli}) using different definitions
of $\meff$ (red and green points correspond to eq.~(\protect\ref{eq:meff}), while blue points to the result of the global fit), 
from eq.~(\protect\ref{eq:fpi_method}) and from the global fit. 
The filled symbols are the results from correlators with source
fixed to a certain time slice, while the open symbols correspond to the average over displaced sources. These tests have been
done on an ensemble with $m_\mathrm{K} = m_\pi \approx 420$~MeV, $a \approx 0.086$~fm and 8000~MDU.}
\label{fig:fpi}
\end{figure}

In principle the same average could be taken also for $y_0$, the position of the source. However as can be seen from Figure~\ref{fig:fpi}
 (left panel), where we plot $F_\pi^\mathrm{bare}$ obtained from eq.~(\ref{eq:fpi_method}) for several $y_0$, 
the correlation functions computed from
displaced sources are fully correlated in the bulk. Therefore, there is no advantage in averaging
over several $y_0$.
In Figure~\ref{fig:fpi} (right panel) we plot the results of $F_\pi^\mathrm{bare}$ (after the plateau average) 
computed with different approaches. All the methods investigated here return 
results in agreement with each other and with the same precision.
Since the global fit is not more precise than the other two methods 
and the one proposed in Ref.~\cite{Guagnelli:1999zf} is sensitive
to the choice of the effective mass used in the exponential factor 
of eq.~(\ref{eq:fpi_guagnelli}), we prefer to use eq.~(\ref{eq:fpi_method}) 
because it does not suffer from the ambiguity of the choice of $\meff$ 
and it turns out to be much easier to implement.

\section{Impact of reweighting on observables}
Simulations with a small pion mass are not only expensive in terms of computational cost but can also run into instabilities 
if accidental zero-modes of the Wilson Dirac operator occur. However such a problem can be cured by regularizing the fermion
determinant with a small twisted-mass term~\cite{Luscher:2008tw}. 
Our set of ensembles has been generated with the so-called
type II twisted-mass reweighting 
(applied only to the Schur complement $\hat Q$ of the asymmetric even-odd preconditioning~\cite{Luscher:2012av}), 
whose fluctuations are smaller in the UV-regime
\begin{equation}
S_\mathrm{f} = -\log \det \frac{(\hat Q^2 + \mu^2)^2}{\hat Q^2+2\mu^2} -2 \log \det Q_\mathrm{oo}\,,
\quad \hat Q = \gamma_5 \hat D \,, 
\label{eq:TM_typeII}
\end{equation}
which requires the computation of the corresponding reweighting factor $W$ to obtain 
the observables in the underlying theory ($\mu=0$, $S_\mathrm{f}=-\log \det \hat Q^2 -2 \log \det Q_\mathrm{oo}$)
\begin{equation}
W = \det \frac{\hat Q^2 (\hat Q^2 + 2 \mu^2)}{(\hat Q^2 + \mu)^2} \,.
\label{eq:rew_factor}
\end{equation}

\begin{figure}
\includegraphics[width=\textwidth]{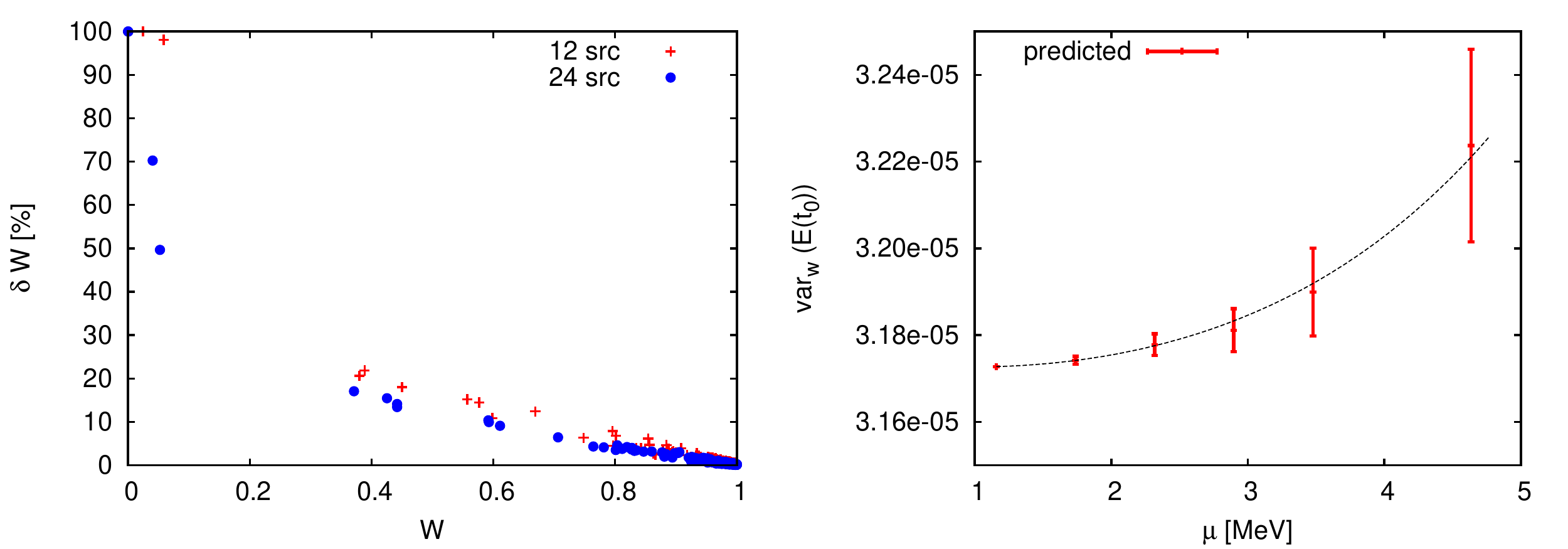}
\caption{\textit{Left}: Error of the reweighting factor as a function of its central value configuration by configuration
for an ensemble with $m_\pi\approx280$~MeV, $a\approx0.086$~fm, 8000~MDU and $\mu=10^{-3}$. Increasing the number of stochastic sources
improves the precision only if $W \gtrsim 0.5$. 
\textit{Right}: the dotted line is drawn only to guide the eye. 
The points are obtained by using the reweighting factor of eq.~(\protect\ref{eq:rew_factor}) 
in eq.~(\protect\ref{eq:var_w})
with increasing $\mu$, 
computed always on the same ensemble (generated with $\mu=0.5\cdot 10^{-3}$, $a\approx 0.086$~fm, $m_\pi \approx 280$~MeV).}
\label{fig:rew-fact}
\end{figure}

The reweighting factors are estimated stochastically, therefore, at first place, it is important  to find an optimal number of
stochastic sources which balances between computational cost and error size, but also to check the convergence of the
variance. Indeed, as can be seen from Figure~\ref{fig:rew-fact} (left panel), 
the relative error of the reweighting
factor grows as the mean value of $W$ goes to zero. In general when $W$ is below 0.8 the stochastic error is larger than
10\% and starts to become relevant, but when $W$ is relatively small, say below 0.5, the computation of the reweighting factor
becomes problematic and even wrong for few gauge-field configurations~\cite{Hasenfratz:2002ym,Finkenrath:2013soa}.
Using a factorization of the determinant 
in eq.~(\ref{eq:rew_factor}) \`a la Hasenbusch
\cite{Hasenbusch:2001ne, Hasenfratz:2008fg} might help in these cases.

A second issue which we investigate here is the possibility 
that the fluctuations of the reweighting factors
with the gauge-field configurations 
influence the precision of the reweighted observables.
Clearly these fluctuations are controlled by the choice of $\mu$, for a fixed light quark mass.
In general $\mu>0$ makes the regions of field space with near-zero modes of the Dirac operator accessible, but large values of
$\mu$ tend to enhance this effect too much. For these configurations $W$ is close to zero while fermionic observables, 
such as meson correlation
functions, develop ``spikes'' and during the reweighting procedure cancellations take place.
Therefore if from one side the twisted mass improves the ergodicity of the algorithm, 
from the other it is a delicate parameter, because it controls the occurrence 
of configurations with almost-zero fermionic
determinant, for which the reweighting procedure might fail (especially without the determinant factorization). 

\begin{figure}
\includegraphics[width=\textwidth, trim=0 3cm 0 0]{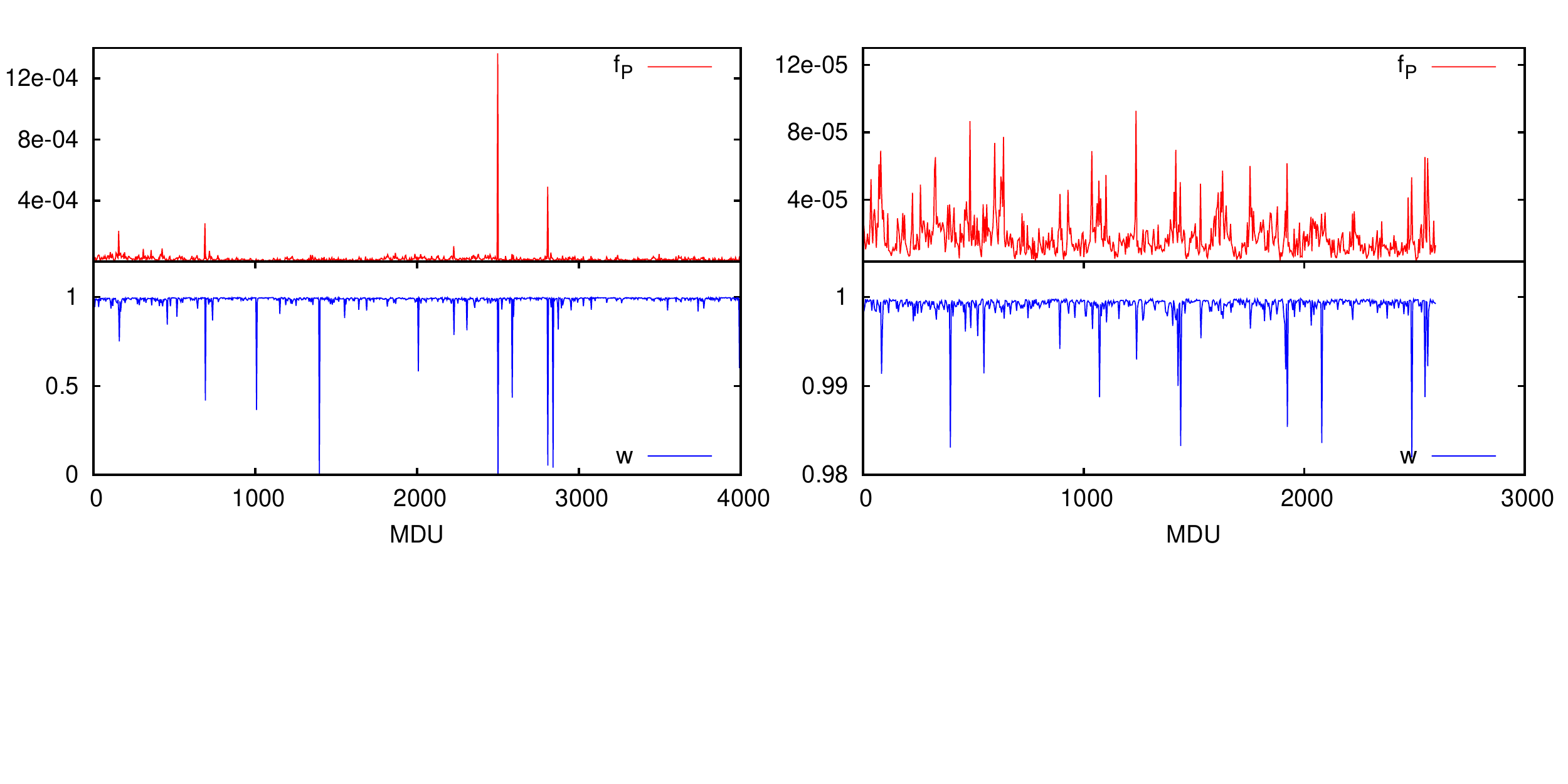}
\caption{The red curves represent the Monte Carlo history of $f_\mathrm{P}(x_0,y_0)$ with $x_0=T/2$, $y_0=a$, the blue curves
show the fluctuations of the reweighting factor from eq.~(\protect\ref{eq:rew_factor}). Both lattices are $96 \times 32^3$ 
and have $m_\pi \approx280$~MeV, $a\approx0.086$~fm. 
\textit{Left}: ensemble generated with $\mu=10^{-3}$. \textit{Right}: ensemble generated with $\mu=0.5 \cdot 10^{-3}$.
Note the difference in the scales of the vertical axes between left and right.}
\label{fig:2pf_rew_H105}
\end{figure}

In Figure~\ref{fig:2pf_rew_H105} we present the Monte Carlo histories of
the twisted-mass reweighting factor and $f_\mathrm{P}(x_0,y_0)$
for two simulations which differ only by the choice of $\mu$.
The fluctuations of both the observable and $W$ 
are suppressed by a factor 10 when $\mu\to\mu/2$ (left$\to$right), 
for this particular setup.
This example illustrates that $\mu$ must be chosen with care.

Since a priori an optimal choice of $\mu$ is unclear,
especially in pioneering simulations like ours, 
it is necessary to study in general how the variance of the reweighted observable is affected
 by the fluctuations of the weights. Given a generic weight function $w$ and observable $O$
\begin{equation}
\langle O \rangle = \frac{ \langle O w \rangle_w}{\langle w \rangle_w} \,, \quad \mbox{with} \quad
\langle O \rangle_w =\frac{ \int dU w^{-1} e^{-S} O}{\int dU w^{-1} e^{-S}} \,,
\label{eq:rew_observable}
\end{equation}
the variance of the reweighted observable depends on $w$ according to the following equation
\begin{equation}
\mathrm{var}_w (O) = \langle w^{-1} \rangle \langle (O-\bar{O})^2 w \rangle \,.
\label{eq:var_w}
\end{equation}
In Figure~\ref{fig:rew-fact} (right panel) we plot eq.~(\ref{eq:var_w}) 
with $O$ being the energy density measured at flow time $t=t_0$ and $W$ the reweighting factor
 as a function of $\mu$.
As a gluonic observable its fluctuations are
expected to be little correlated to those of the reweighting factor.
Hence $\mathrm{var}_w (E(t_0))$ is largely 
independent of the choice of the twisted-mass regulator
and this is confirmed by a negligible increase 
of the variance in Figure~\ref{fig:rew-fact} with $\mu$. 
The situation is different for mesonic correlators and needs more investigation.

\section{Conclusions}
In these Proceedings we have discussed the effects induced by the open boundaries
on some gluonic and fermionic observables. They are dominated by discretization errors
which however decrease rapidly with the lattice spacing.
We have shown that vacuum expectation values of scalar quantities, such as $t_0$ or the
meson masses, can be taken in a safe region in the bulk of
the lattice. We have analysed different strategies to extract pseudoscalar decay constants
 (in the presence of boundaries) and
we have also proposed a new method.
At the moment, none seems to be preferable over the others.
The breaking of translation invariance in time 
given by the boundaries does not introduce a significant
disadvantage in terms of precision and spectral 
quantities can be computed with sub-percent accuracy.
In particular source displacement does not reduce the final statistical error.

We have studied the effects of a twisted-mass regulator in the light sector.
Also with the reweighting, we are able to obtain results
with a small statistical uncertainty.
The total cost or benefit of this method is difficult to assess:
a cheaper simulation on the one hand has to be confronted
with a (slightly) increased uncertainty on the other.
For now, we conclude that the method works well.

\acknowledgments

We thank B.~Leder for collaboration in an early stage of the work.
We would also like to thank D.~Djukanovic, G.~P.~Engel, A.~Francis, G.~Herdoiza, 
H.~Horch, M.~Papinutto, E.~E.~Scholz, J.~Simeth, H.~Simma and W.~S\"oldner for their
effort in the production of the ensembles and for many useful discussions.
We acknowledge PRACE for awarding us access to resource FERMI based in Italy at
CINECA, Bologna and to resource SuperMUC based in Germany at LRZ, Munich.
We had access to HPC resources in the form of
a regular GCS/NIC project$^3$, a JUROPA/NIC project\footnote{
http://www.fz-juelich.de/ias/jsc/EN/Expertise/Supercomputers/ComputingTime/Acknowledgements.html}
and through PRACE-2IP, receiving funding from the
European Community's Seventh Framework Programme (FP7/2007-2013) under grant
agreement RI-283493.
This work is supported in part by the grants SFB/TR9 of the Deutsche Forschungsgemeinschaft.